\documentclass[sn-mathphys]{sn-jnl-modified}

\jyear{2021}%

\theoremstyle{thmstyleone}%
\theoremstyle{thmstyletwo}%

\theoremstyle{thmstylethree}%

\usepackage{mathtools,amssymb}
\usepackage{xcolor}
\usepackage{amsthm}
\usepackage{hypcap}
\usepackage{xfrac}
\usepackage{amssymb}
\usepackage{xfrac}
\usepackage{graphicx}
\usepackage{graphics}
\usepackage{float}
\usepackage{subcaption}
\usepackage{caption}
\usepackage{multirow,multicol, makecell, booktabs}
\usepackage{multirow,multicol, makecell, booktabs}
\usepackage{multirow}
\usepackage{amsmath}
\usepackage{array}
\usepackage{multirow}
\newcolumntype{L}{>{$}l<{$}}
\newcolumntype{C}{>{$}c<{$}}
\usepackage{mathrsfs}
\usepackage{tabularx}
\usepackage{adjustbox}
\usepackage{tabularx}

\newcommand{\vect}[1]{\left( #1_1, ..., #1_N\right)}
\usepackage{hyperref}

\usepackage{academicons}
\usepackage{xcolor}

\newcommand{\orcid}[1]{\href{https://orcid.org/#1}{\textcolor[HTML]{A6CE39}{\aiOrcid}}}

\begin{document}

\title[Combining Multiple Testing with Multivariate Singular Spectrum Analysis]
{Combining Multiple Testing with Multivariate Singular Spectrum Analysis}


\author[1]{\fnm{Maryam} \sur{Movahedifar}}\email{movahedm@uni-bremen.de}

\author*[2]{\fnm{Thorsten} \sur{Dickhaus}  }\email{dickhaus@uni-bremen.de} 

\affil[1,2]{\orgdiv{Institute for Statistics}, \orgname{University of Bremen}, \orgaddress{\street{Bibliothekstraße1}, \city{Bremen}, \postcode{28359}, \country{Germany}}}
\affil[1]{\small {ORCID:} 0000-0003-0283-7560}
\affil[2]{\small {ORCID:} 0000-0003-3084-3036}


\abstract{Appropriate preprocessing is a fundamental prerequisite for analyzing a noisy dataset. The purpose of this paper is to apply a nonparametric preprocessing method, called Singular Spectrum Analysis (SSA), to a variety of datasets which are subsequently analyzed by means of multiple statistical hypothesis tests.
SSA is a nonparametric preprocessing method which has recently been utilized in the context of many life science problems. In the present work, SSA is compared with three other state-of-the-art preprocessing methods in terms of goodness of denoising and in terms of the statistical power of the subsequent multiple test. These other methods are either parametric or nonparametric. Our findings demonstrate that (multivariate) SSA can be taken into account as a promising method to reduce noise, to extract the main signal from noisy data, and to detect statistically significant signal components.}




\maketitle

\section{Introduction}\label{section.Intro}
Most real-life datasets are contaminated by the presence of considerable noise. Thus, noise reduction and signal extraction have always been considered as important steps of data analysis in all fields of study; see, among many others, \cite{Pyle1999,Repsilber2008,OLIVERI20199}. In this paper, different types of preprocessing methods, both parametric and nonparametric, are evaluated. 
In particular, we consider the Autoregressive Fractionally Integrated Moving Average (ARFIMA) model (see \cite{Hyndman20081}) as well as Exponential Smoothing (ETS, see \cite{b1}) as parametric methods, and Singular Spectrum Analysis (SSA, cf. Section \ref{section.ssa} for details) and feed-forward Neural Network (NN, see \cite{Bishop-NN,b1}) as nonparametric methods.
The usage of nonparametric models is often considered when parametric model assumptions are prone to be violated (cf., for instance, \cite{Dickhaus2018,Hassani2013}). One example is constituted by economic time series: Economic conditions (in most cases, especially after recessions) can lead to non-constant mean and variance, thus to a violation of parametric assumptions of stationarity. 

The first purpose of the present paper is to compare the aforementioned signal preprocessing techniques in terms of their accuracy in signal extraction by means of simulated (time series) data where ground truth is known. The second purpose is to provide a comparison on real data. In the latter case, ground truth is not known, and we compare the methods in terms of a penalized (empirical) signal-to-noise ratio. Furthermore, we compare them in terms of the number of rejected null hypotheses of a variety of multiple tests (see \cite{Handbook,thorsten} for comprehensive introductions to multiple testing) in the case that the aforementioned preprocessing methods are applied to real data prior to carrying out the multiple test. In all our comparisons, (multivariate) SSA turns out to be a promising approach.

The rest of this paper is organized as follows. Section \ref{section.ssa}  provides a short theoretical background of basic SSA. Section \ref{section.methods} describes the other three considered preprocessing methods. In Section \ref{sim result}, we report results from a simulation study comparing the signal extraction accuracies of the considered preprocessing methods. Section \ref{section.mt}  is devoted to a brief introduction to multiple testing. In Section \ref{section.real data}, several real-life datasets are analyzed. Finally, Section \ref{section.conclusion} contains a discussion.

\section{Singular Spectrum Analysis}\label{section.ssa}
The origins of SSA can be traced back to the work of Broomhead and King (1986); see \cite{BROOMHEAD1986217}. SSA, as a branch of time series analysis, is a nonparametric approach that is free from the necessity of  statistical assumptions like stationary or linearity of the time series, which are unlikely to hold in the real world. The SSA method is also known for its ability to deal with short time series in which classical methods fail due to a lack of a sufficient number of observations. Furthermore, the SSA method is capable of filtering and forecasting time series. It can be carried out as a univariate or as a multivariate method. 

In short, the SSA technique initially filters the time series to decrease the noise, and then reconstructs the less noisy series to forecast the time series. The major goal of the SSA method is  to decompose the initial ordered series (e.\ g., a time series) into a sum of  different components which can be considered as either a trend, a periodic, a quasi-periodic (perhaps, amplitude-modulated), or a noise element. We refer to 
\cite{MOVAHEDIFAR201852,  SILVA2019134, article35, HASSANI38, article39, signals} for more details.

As mentioned before, SSA (henceforth used to refer to the univariate version) is currently widely used, with  a considerable number of applications. However, compared to SSA, multivariate SSA (MSSA) is still in its initial stage and only a handful of applications of MSSA  have been investigated so far; see, for example, \cite{SILVA2017,DECARVALHO,articlehs,articlepl}. In the following subsections, concise descriptions of SSA and MSSA are presented. For more information, see \cite{Golyandina-2001,Golyandina-Zhigljavsky-2013,Golyandina-R-2018}.

\subsection{A Brief Description of Basic SSA}\label{section.intssa}
Let $\textbf{Y}_N=\vect{y}$ denote a time series of length \textit{N}. Fix an integer $\textit{L}$, which is called window length, where ~$2\leq \textit{L} \leq N/2$. Then, basic SSA consists of the following four steps:
\begin{description}
	\item [1) \textbf{ Embedding}:]
	This step transfers the one-dimensional time series $\textbf{Y}_N=\vect{y}$ of length $N$ into the multi-dimensional series $\mathbf{X}^{(1)}, \ldots, \mathbf{X}^{(K)}$ with vectors $\textbf{X}^{(i)} =\left(y_i,\ldots, y_{i+L-1}\right)^{T}\in\mathbb{R}^{L}$, where $K = N - L +1$. The vectors $\mathbf{X}^{(i)}$ are called $L$-lagged vectors (or, simply, lagged vectors). The embedding step includes only one parameter, namely, the window length 
	$L$. The purpose of this step is to form the trajectory matrix $\textbf{X} =\left[\mathbf{X}^{(1)},\ldots, \mathbf{X}^{(K)}\right] \in \mathbb{R}^{L \times K}$.
	
	\item[ 2) \textbf{Singular Value Decomposition (SVD)}:]
	In this step, the trajectory matrix $\mathbf{X}$ is decomposed into a sum of
	rank-one elementary matrices. Let $\lambda_1,\ldots,\lambda_L$ denote the eigenvalues of
	$\mathbf{X}{\mathbf{X}}^{T}$, which are ordered 
	 in decreasing magnitude such that
	$\lambda_1\geq\dots\geq\lambda_L\geq 0$, and let
	$\mathbf{U}_1, \ldots, \mathbf{U}_L$ denote the eigenvectors of the matrix
	$\mathbf{X}{\mathbf{X}}^{T}$ corresponding to these eigenvalues.  Letting
	$d = \textrm {max} \{i: \lambda_i> 0 \} = \text{rank}(\textbf {X}) $, 
	the SVD of the trajectory matrix can be written as $\mathbf{X}=\mathbf{X}_1+\cdots+\mathbf{X}_d $, where  $\textbf {X}_i ={\sqrt\lambda_i} \mathbf{U}_i {\mathbf{V}_i}^{T}$ and $\mathbf{V}_i=\textbf {X}^{T} \mathbf{U}_i/{\sqrt\lambda_i}$ for $i=1,\ldots,d$.
	
	\item[3) \textbf{Grouping}:]
	In this step, the set of indices $\{1,\ldots,d\}$ is partitioned into $\gamma$ disjoint subsets ${I}_1,\ldots, {I}_\gamma$. Let ${I}=\{i_1,\ldots,i_p\}$. Then, the matrix $\mathbf{X}_I$ corresponding to the group $I$ is given by $\mathbf{X}_I=\mathbf{X}_{i_1}+\cdots+\mathbf{X}_{i_p}$.
	For example, if ${I}=\{2,3,5\}$, then $\mathbf{X}_I=\mathbf{X}_{2}+\mathbf{X}_{3}+\mathbf{X}_{5}$. Considering the SVD of $\mathbf{X}$, the split of the set of indices $\{1,\ldots,d\}$ into the disjoint subsets
	$I_1,\ldots,I_\gamma$ can be represented as $ \mathbf{X}=\mathbf{X}_{I_1}+\cdots+\mathbf{X}_{I_\gamma}
	 $. The choice of the groups is explained, for instance, in Section 2.1.2.3 of \cite{Golyandina-Zhigljavsky-2013}.
	
	\item[4) \textbf{Diagonal Averaging}:] The aim of this step is to transform every obtained matrix $\mathbf{X}_{I_j}$ from the grouping step into a Hankel matrix so that these can subsequently be converted back into a (reconstructed) time series. In basic SSA, Hankelization is achieved via diagonal averaging of the matrix elements over the anti-diagonals, see \cite{Golyandina-2001}.
\end{description}

\subsection{Multivariate \textrm{SSA}}
In the case that multiple time series are observed, the joint structure or the dependency among the series are also of interest, in addition to the internal structure of each individual time series. This is incorporated in MSSA. 
The main idea of the MSSA method is similar to that of basic SSA, but the embedding step for constructing the trajectory matrix is different. The purpose of MSSA is to take into consideration the combined structure of a multivariate series to exploit information about the joint distribution of these series. 

Here we consider the algorithm of MSSA for analyzing multivariate time series which is explained in Section 4.2 of \cite{Golyandina-R-2018}. A brief description of the MSSA method is as follows:  
Suppose a multivariate time series is given as a collection $\{{\textbf{Y}}^{(p)}=({y_j}^{p})_{j=1}^{N_p}: p=1,\ldots,s\}$ of $s$ univariate time series of length $\textit{N}_p$ for each $p \in \{1,\ldots,s\}$. The generic scheme of the algorithm described in Section \ref{section.intssa} is also used in MSSA; it is just required to describe the embedding step for creating the trajectory matrix $\textbf{X}_{\text{MSSA}}$. To this end, let the integer $L$ again denote the window length, where $2\leq L < \min(N_p: p=1,\ldots,s)$. For each time series $\mathbf{Y}^{(p)}$, we form  $L$-lagged vectors ${\mathbf{X}_j}^{(p)}=\left(y_j^p,\ldots, y_{j+L-1}^p\right)^T$ for $1\leq j \leq K_p$, where $K_p = N_p - L +1$. Letting $K=\sum_{p=1}^{s}{K_p}$, the trajectory matrix of $\{{\textbf{Y}}^{(p)}=({y_j}^{p})_{j=1}^{N_p}: p=1,\ldots,s\}$ is then a matrix of size $L\times K$ and has the form
\(
\textbf{X}_{\text{MSSA}}=[{\textbf{X}_1}^{(1)}, \ldots, {\textbf{X}_{K_1}}^{(1)}, \ldots, {\textbf{X}_1}^{(s)}, \ldots, {\textbf{X}_{K_s}}^{(s)}] 
\).\\
 For more details on MSSA and its properties we refer to Chapter 4 in \cite{Golyandina-R-2018}.

\section{Other Filtering Methods}\label{section.methods}
In this section, the three other considered (parametric and nonparametric) signal extraction methods are briefly explained. 

\subsection{ARFIMA}\label{section.arfima}
The autoregressive fractionally integrated moving average (ARFIMA) model of order $(p, d, q)$, 
denoted by ARFIMA$(p, d, q)$, is useful in describing phenomena of long-memory processes. The ARFIMA model is also a generalization of the autoregressive integrated moving-average (ARIMA) model with integer degrees of integration. ARFIMA models are appropriate for overdifference stationary series that display long-run dependence. In the ARIMA process, a nonstationary time series is differenced $d \in \mathbb{N}$ times, with the goal of achieving stationarity of the differenced time series. These kinds of time series are said to be integrated of order $d$, denoted $I(d)$. Stationary series correspond to not differencing, hence, they are $I(0)$. Some time series exhibit too much auto-dependence to be $I(0)$, but are not $I(1)$, either. The ARFIMA model allows for a continuum of fractional differences $d \in (-0.5, 0.5)$. The generalization to fractional differences allows the ARFIMA model to handle processes that are neither $I(0)$ nor $I(1)$, to test for overdifferencing, and to model long-run effects that only die out at long time horizons; see \cite{arfima}. 

Formally, the ARFIMA model with mean $\mu$ has the specification
\[
\Phi (L) ({1-L})^{d}(y_t - \mu)=\theta (L)\varepsilon_t,
\] 
where the $\varepsilon_t$'s are stochastically independent and identically distributed (i.i.d.) with zero mean and variance $\sigma_{\varepsilon}^2$, $L y_t = y_{t-1}$, $\Phi (L)= 1- \Phi_{1} L- \ldots -\Phi_{p} L^p$, $\theta (L)= 1+ \varphi_{1} L+ \ldots +\varphi_{q} L^q$, and $({1-L})^{d}$ is equal to
\[
({1-L})^{d}= \sum_{k=0}^{\infty}{d \choose k }(-L)^{k}.
\]
The real numbers $\Phi_1, \ldots, \Phi_p$ and $\varphi_{1}, \ldots, \varphi_{q}$ are model parameters. The ARFIMA model can be fitted by minimizing the Akaike Information Criterion (AIC). For more details, see \cite{arfima2}.

\subsection{ETS}\label{section.ets}
The Exponential Smoothing (ETS) method builds upon a state space model for the time series under consideration. For example, in Section 3 of \cite{HYNDMAN2002439} the following system of equations is considered:
\begin{eqnarray*}
Y_t &=& h(\mathbf{x}_{t-1}) + k(\mathbf{x}_{t-1}) \varepsilon_t,\\
\mathbf{x}_t &=& f(\mathbf{x}_{t-1}) + g(\mathbf{x}_{t-1}) \varepsilon_t.
\end{eqnarray*}
Different choices of the vectors $(\mathbf{x}_t)_t$ and the functions $f$, $g$, $h$, and $k$ lead to different types of (parametric) models; see Table 1 in 
\cite{HYNDMAN2002439}. These vectors and functions are capable of expressing different time series components such as, e.\ g., trend and season. The term ETS itself refers to a weighting scheme that is used in forecasting future time series values, where the weights decrease exponentially with growing lag.

It is beyond the scope of the present paper to describe the ETS method comprehensively, and we refer to \cite{Hyndman2008} and Chapter 8 of \cite{b1} instead. For our applications, we have used the ETS model from the \texttt{R} package \texttt{forecast}; see 
\url{https://cran.r-project.org/web/packages/forecast/forecast.pdf}.  In this implementation, model selection is performed by minimizing the AIC, and model parameters are estimated by using the maximum likelihood approach. 

\subsection{Neural Network}\label{section.nn}
The Neural Network (NN) model that we have considered takes the form
\begin{equation}\label{nnetar1}
\hat{y}_{t}=\hat{\beta}_{0}+\sum_{j=1}^{k}\hat{\beta}_{j}\psi(\mathbf{x}_{t}, \hat{\gamma}_{j}),
\end{equation}
where $\mathbf{x}_{t} \in \mathbb{R}^p$ consists of $p$ lagged values of $y_{t}$. The functional form of $\psi$ is typically sigmoidal. For instance, the logistic form of $\psi$ is given by
\begin{equation}\label{eq.nn}
\psi(\mathbf{x}_{t}, \hat{\gamma}_{j})={\left[1+ \exp\left(-\hat{\gamma}_{j0} - \sum_{i=1}^{p}\hat{\gamma}_{ji} \cdot y_{t-i}\right)\right]}^{-1}, \; j=1, \ldots, k.
\end{equation}
Equations \eqref{nnetar1} and \eqref{eq.nn} define a feed-forward neural network with a single hidden layer and lagged time series inputs. The number $k$ of logistic functions used in \eqref{eq.nn} is equal to the number of hidden nodes. Neural networks with a data-dependent choice of $k$ are occasionally referred to as nonparametric neural networks in the literature; see, e.\ g., \cite{Philipp-ICLR2017}. We adopt this terminology here.
In our applications, we have used the nonparametric NN implementation \texttt{nnetar} from the \texttt{R} package \texttt{forecast}; see also Section 12.4 in \cite{b1} for details.

\section{Computer Simulations}\label{sim result}
The purpose of this section is a comparative evaluation of the preprocessing methods described in Sections \ref{section.ssa} and \ref{section.methods} regarding their capability of denoising a dataset. For this purpose, we have generated times series $(y_{t}: t=1, \ldots, N = 100)$ according to a "signal plus noise" model of the form $y_t = f_t + \varepsilon_t$, $t=1, \ldots, N$. For the signal part $(f_t)_{t=1, \ldots, N}$, the following four choices have been considered.
\begin{enumerate}
\item \textbf{Sine+Exponential}\label{e1} : $f_{t}=\sin(2\pi t/12)+\exp(0.01 t)$
\item \textbf{Cosine+Linear}: $f_{t}=0.8\cos(\pi t/3)+0.6t$
\item \textbf{Sine$\times$Exponential}: $f_{t}=\sin(2\pi t/12) \times \exp(0.01 t)$
\item \textbf{Sine+Linear+Exponential}: $f_{t}=\sin(3\pi t/12)+0.5t+\exp(0.03 t)$
\end{enumerate}
In all four cases, the (noise) time series $(\varepsilon_t: t=1, \ldots, N=100)$ consisted of (centered) Gaussian white noise with unit error variance $\sigma^2 = 1$.

Each of the above series has been simulated $1000$ times. The denoising performance of SSA, as compared to each of the methods described in Section \ref{section.methods}, has been assessed by calculating the relative root mean squared error (RRMSE) and the relative mean absolute error (RMAE) criteria, which are given by 
\begin{eqnarray*}
	\textmd{RRMSE} &= &\frac{\left(\sum_{i=1}^{N}(f_{i}-\hat{f}_{i})^{2}\right)^{1/2}}{\left(\sum_{i=1}^{N}(f_{i}-\tilde{f}_{i})^{2}\right)^{1/2}},\\ 
	\textmd{RMAE} &= &\frac{\sum_{i=1}^{N} \lvert f_{i}-\hat{f}_{i} \rvert }{\sum_{i=1}^{N} \lvert f_{i}-\tilde{f}_{i} \rvert},
\end{eqnarray*}
where $\hat{f}_{i}$ is the estimated value of $f_{i}$ based on SSA and $\tilde{f}_{i}$ is the estimated value of $f_{i}$ obtained by applying a competing method. For each of the methods that we have discussed in Section \ref{section.methods}, the values $(\tilde{f}_{i})_{1 \leq i \leq N}$ have been obtained by using the \texttt{R} function \texttt{fitted()}. Essentially, this function first computes the residuals of the model by replacing unknown quantities by their estimates in the model equation. Then, these residuals are subtracted from the original data points. If $\text{RRMSE} < 1$ or $\text{RMAE} < 1$, respectively, then the SSA procedure outperforms the other method by $(1- \text{RRMSE}) \times 100\%$ or 
$(1- \text{RMAE}) \times 100\%$, respectively.

Table \ref{tab.simulation} reports the average RRMSE and RMAE values (averaged over the $1000$ Monte Carlo repetitions of the simulation) attained under each of the four aforementioned simulation 
models. It becomes apparent from Table \ref{tab.simulation} that the SSA method has a higher accuracy in signal extraction than the other considered preprocessing methods, for all four considered data-generating models and for all considered choices of the tuning parameter $L$ (window length).

\begin{sidewaystable}
\sidewaystablefn%
\begin{center}
\begin{minipage}{\textheight}
\caption{RRMSE and RMAE values rounded to two decimal places for reconstructing the signal underlying four simulated time series}\label{tab.simulation}
\begin{tabular*}{\textheight}{@{\extracolsep{\fill}}llccccccccc@{\extracolsep{\fill}}}
\toprule%
&&\multicolumn{3}{@{}c@{}}{SSA/ARFIMA}& \multicolumn{3}{@{}c@{}}{SSA/ETS}& \multicolumn{3}{@{}c@{}}{SSA/NN}
\\\cmidrule{3-5}\cmidrule{6-8}\cmidrule{9-11}%
Simulation Series & Criterion & $L\footnotemark[1]=10$	& $L=30$ & $L=50$ & $L=10$	& $L=30$ & $L=50$& $L=10$	
& $L=30$ & $L=50$ \\
\midrule
\multirow{2}{*}{Sine+Exponential} &{RRMSE\footnotemark[2]}& 0.43 & 0.32& 0.20 &0.40 &0.20 & 0.14 &0.71& 0.30& 0.20\\
&{RMAE\footnotemark[3]}&0.42& 0.33  &0.19 &0.40  &0.20 &0.12 &0.72 &0.30& 0.18\\\hline
\multirow{2}{*} {Cosine+Linear}   &{RRMSE} & 0.22 & 0.14 & 0.13 & 1.07 & 0.67& 0.65 & 0.57& 0.36& 0.35 \\
&{RMAE} & 0.25 & 0.16 & 0.15 & 0.96 & 0.60 & 0.58& 0.57& 0.36& 0.34\\\hline
\multirow{2}{*}{Sine$\times$Exponential} &{RRMSE} & 0.45& 0.28& 0.27&0.34& 0.21& 0.20& 0.50& 0.32& 0.31 \\
&{RMAE} &0.44& 0.28& 0.27& 0.33& 0.21& 0.20& 0.51& 0.33& 0.31\\\hline
\multirow{2}{*}{Sine+Linear+Exponential} &{RRMSE} & 0.21& 0.13& 0.12& 0.73& 0.45& 0.44& 0.58& 0.35& 0.35  \\
&{RMAE} &0.24& 0.15& 0.14& 0.69& 0.42& 0.41& 0.58& 0.35& 0.34\\
\botrule
\end{tabular*}
\footnotetext[1]{Window Length ($L$)}
\footnotetext[2]{Relative Root Mean Squared Error}
\footnotetext[3]{Relative Mean Absolute Error}
\end{minipage}
\end{center}
\end{sidewaystable}

\section{Multiple Hypotheses Testing}\label{section.mt}
In this section, we explain some fundamental concepts regarding multiple test problems and multiple test procedures (MTPs). In general, a multiple test problem occurs whenever $m > 1$ null hypotheses $H_1, \ldots, H_m$ shall be tested simultaneously based on one and the same dataset. Denoting the sample space of the statistical model under consideration by $\mathcal{Y}$, a (non-randomized) multiple test for $H_1, \ldots, H_m$ is a (measurable) mapping $\varphi=(\varphi_1, \ldots, \varphi_m)^{T}: \mathcal{Y}\to \{0,1\}^m$ the components of which have the usual interpretation of a statistical test for $H_i$ versus $K_i$, $1 \leq i \leq m$. Namely, $H_i$ is (by convention) rejected if and only if $\varphi_i(\mathbf{y})=1$, where $\mathbf{y} \in \mathcal{Y}$ denotes the observed data.

Many multiple tests operate on test statistics $T_1, \ldots, T_m$ or corresponding $p$-values $p_1, \ldots, p_m$, respectively. Utilizing $p$-values is useful in the multiple testing context, because every $p$-value takes its values in the unit interval $[0,1]$, even if the corresponding test statistics have different scales. Thus, transforming test statistics into $p$-values serves the purpose of a standardization.  It is beyond the scope of the present paper to discuss the notion of $p$-values in depth, and we refer to Chapter 2 in \cite{thorsten} for details. For our purposes, it suffices to know that a (valid) $p$-value, regarded as random variable, is under the null hypothesis stochastically not smaller than the uniform distribution on $[0, 1]$, and that $p$-values typically tend to small values under alternatives.

\subsection{Error Control of a Multiple Test}\label{section.mt.def}
Assuming that exactly $m_0$ of the $m$ null hypotheses are true and the remaining $m_1 = m - m_0$ ones are false (where $m_0$ and $m_1$ are unknown in practice), the decision structure of a given multiple test is given in Table \ref{tab.test1}. The random variable $V_m$ counts the number of type I errors, and the random variable $T_m$ counts the number of type II errors. Unlike in one-dimensional test problems (i.\ e., in cases with $m=1$), type I and type II errors can occur simultaneously if $m > 1$. The total number of rejections
is denoted by $R_m$ in Table \ref{tab.test1}. Notice that only $m$ and $R_m$ are observable in practice, because all the quantities $U_m$, $V_m$, $T_m$, and $S_m$ dependent on the unknown value of the parameter $\theta$ (say) of the statistical model under consideration.
\begin{table}[htp]
\begin{center}
\begin{minipage}{125pt}
\caption{Decision structure of a multiple test procedure}\label{tab.test1}%
\begin{tabular}{@{}llll@{}}
\toprule
  \multicolumn{4}{c}{Test decisions} \\\hline
  Hypotheses & 0 & 1 \\
  True & $U_m$&$V_m$& $m_0$  \\
  False & $T_m$ & $S_m$& $m_1$ \\
    &$W_m$&$R_m$&$m$ \\
\botrule
\end{tabular}
\end{minipage}
\end{center}
\end{table}
\noindent
Based on the quantities given in Table \ref{tab.test1}, a variety of multiple type I and type II error rates have been introduced in the literature. The traditional type I error criterion for a multiple test, which we will also apply in Section \ref{section.real data}, is (strong) control of the family-wise error rate (FWER). The number $\text{FWER}_{\theta}(\varphi)=\mathbb{P_{\theta}}(V_m(\varphi) > 0)$ is called the FWER of the multiple test $\varphi$ under $\theta$, and $\varphi$ is said to control the FWER strongly at level $\alpha \in (0,1)$, if $\sup_{\theta\in \Theta} \text{FWER}_{\theta}(\varphi)\leq \alpha$, where $\Theta$ is the parameter space of the statistical model under consideration. The power  of a given multiple test is typically defined in terms of $S_m$, which is the (random) number of correctly rejected, false null hypotheses. In Section \ref{section.real data}, we will compare concurring multiple tests by means of $R_m$. If it is guaranteed that the concurring multiple tests all control the FWER at the same level $\alpha$, a higher number of rejections is an indication for higher power.

\subsection{FWER-Controlling Multiple Tests}\label{multiple-tests-list}
In Section \ref{section.real data}, we will apply the following multiple tests.
\begin{itemize}
\item[(a)] Bonferroni correction, cf. \cite{Bonferroni1,Bonferroni2}: Each individual test $\varphi_i$ is carried out at the "local" significance level $\alpha / m$. Due to the first-order Bonferroni inequality (also known as Boole's inequality or the union bound), this type of multiplicity correction yields a strongly FWER-controlling multiple test.
\item[(b)] Holm's procedure, cf. \cite{holm}: Holm's procedure is a step-down variant of the Bonferroni method. The smallest $p$-value is compared with $\alpha/m$. If it is smaller than this threshold, the corresponding null hypothesis is rejected and the second smallest $p$-value is compared with $\alpha / (m-1)$, and so on. The procedure stops as soon as one of the ordered $p$-values is not smaller than its threshold, and the corresponding null hypothesis as well as all null hypotheses corresponding to larger $p$-values are retained. Holm's procedure controls the FWER strongly.
\item[(c)] \v{S}id\'{a}k correction, cf. \cite{sidak1967}: Each individual test $\varphi_i$ is carried out at the local significance level $1 - (1 - \alpha)^{1/m}$. This type of multiplicity correction yields a strongly FWER-controlling multiple test if the (random) $p$-values $p_1, \ldots, p_m$ are jointly stochastically independent, or if they exhibit certain forms of positive dependence; see \cite{MPHTP} for mathematical details.
\item[(d)] \v{S}id\'{a}k step-down: Step-down variant of the \v{S}id\'{a}k procedure. It controls the FWER strongly if the (random) $p$-values $p_1, \ldots, p_m$ are jointly stochastically independent, or if they exhibit certain forms of positive dependence.
\item[(e)] Hochberg's procedure, cf. \cite{hoch}: Hochberg's procedure is a step-up test. The largest $p$-value is compared with the threshold $\alpha = \alpha / 1$. If it is smaller than this threshold, then all $m$ null hypotheses are rejected. If it is not smaller than $\alpha$, then the corresponding null hypothesis is retained and the second largest $p$-value is compared with the threshold $\alpha / 2$, and so on. Hochberg's procedure controls the FWER strongly, if the Simes inequality (cf. \cite{sim}) holds for $p_1, \ldots, p_m$. Conditions for the latter have been discussed, for example, in \cite{Bodnar-Dickhaus-AISM} and 
\cite{Finner-Simes}. Under joint independence of $p_1, \ldots, p_m$,  the Simes inequality is guaranteed to hold, but certain forms of (positive) dependence among $p_1, \ldots, p_m$ also lead to the validity of the Simes inequality.
\end{itemize}

For applications of multiple testing, especially in the life sciences, see 
\cite{thor1,thor2,thor3,thor4,dickhaus2}, and Parts II and III in \cite{thorsten}. 

\section{Analyzing Real Data}\label{section.real data}
\subsection{Overview of the Datasets}\label{section.overview.data}
\begin{figure}[htp]
\centering
\includegraphics[width=0.9\textwidth]{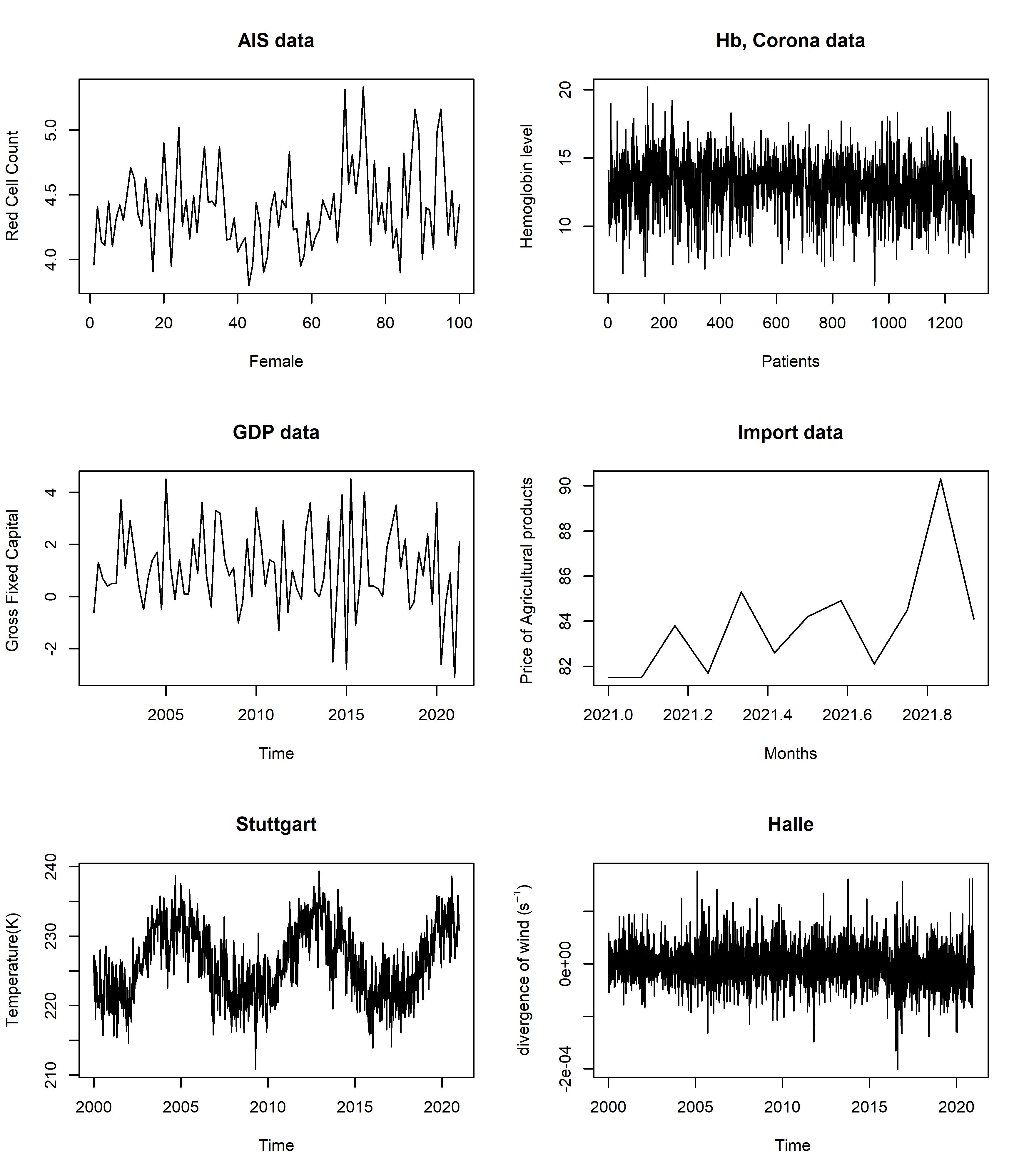}
\caption{Six examples of noisy datasets}\label{noisy data}
\end{figure}

For illustrative purposes, Figure \ref{noisy data} contains plots of selected noisy data series taken from the real data examples which we will investigate deeper with respect to multiple test problems in Sections \ref{sport.data} - \ref{weather.data} below.

The left graph in the upper row of Figure \ref{noisy data} displays data from Section \ref{sport.data}, representing the red blood cell counts of $100$ female athletes.  The right graph in the upper row of Figure \ref{noisy data} displays data from Section \ref{corona.data}, representing the hemoglobin levels of $1302$ Coronavirus patients. The left graph in the middle row of Figure \ref{noisy data} displays data from Section \ref{GDP.data}, representing the time series of gross fixed capital in the United Kingdom (UK) over recent years. The right graph in the middle row of Figure \ref{noisy data} displays data from Section \ref{ie.data}, representing the time series of agricultural products prices for import data in Germany during the year 2021. Finally, the lower row in  Figure \ref{noisy data} displays data from Section \ref{weather.data}, representing climate time series of two German cities, namely, a temperature time series of Stuttgart (left) and a divergence of wind time series of Halle (right), from the year 2000 until present.

Despite originating from different scientific areas and being of different structure, all six graphs displayed in Figure \ref{noisy data} exhibit a unifying feature, namely, the presence of pronounced noise, leading to highly volatile patterns in the six graphs. Therefore, appropriate preprocessing of these datasets appears necessary prior to a statistical analysis. 

\subsection{Assessing the Goodness of Denoising on Real Data}
Due to the fact that the signal component $f$ in the analysis of real data is unknown, the criteria defined in Section \ref{sim result} cannot be used to evaluate methods in the real-data scenario. Therefore, we need another criterion to compare their results, cf. \cite{ghafari}. The most commonly used criterion for evaluating a denoising method in a real dataset is the (empirical) signal-to-noise ratio (SNR), which has different definitions according to the applications. The SNR criterion calculates the relative magnitude of the received signal power relative to the noise. SNR is usually measured in the units of decibels (dB), and since the  units of signal power and noise are the same, the formula works for any kinds of units.  Here, we apply the following formula for the (empirical) signal-to-noise ratio:
\begin{equation}\label{SNR-def}
\text{SNR}(\hat{\mathbf{f}})= \frac{\sum_{i=1}^{N}\hat{f}_{i}^{2}}{\widehat{\text{Var}}(\mathbf{y}-\hat{\mathbf{f}})}, 
\end{equation}
where $\mathbf{y}$ are the observed data, $\hat{\mathbf{f}} = (\hat{f}_{1}, \ldots, \hat{f}_{N})^{T}$ denotes the estimated signal values after denoising, and $\widehat{\text{Var}}$ denotes the empirical variance. In principle, larger SNR values indicate the superiority of a method. However, the empirical variance in \eqref{SNR-def} can be made artificially small by the use of a very complex model for $\hat{\mathbf{f}}$. This effect is commonly referred to as overfitting. Therefore, we penalize the (empirical) signal-to-noise ratio by  a criterion that measures the smoothness or roughness, respectively, of a signal. To this end, we utilize the roughness measure mentioned around Equation (5.1.5) in \cite{smooth2}. Namely, the roughness of a function $\mathit{f}$, which has continuous first and second derivatives, is defined by
\begin{equation}\label{rough}
\text{Roughness}(f) = \int (f^{\prime\prime}(u))^{2} du.
\end{equation}
A lower amount of roughness is related to smoother function. As a discrete version of \eqref{rough}, we can define 
\begin{equation}
\text{Roughness}_N(\mathbf{f})=\sum_{n=3}^{N}(\bigtriangledown^{2}f_{n})^{2}=\sum_{n=3}^{N}(\bigtriangledown(\bigtriangledown f_{n}))^{2},
\end{equation}
where $\mathbf{f} = (f_1, \ldots, f_N)^{T}$ and $\bigtriangledown$ is the difference operator, given by $\bigtriangledown f_{n}= f_{n}-f_{n-1}$. In our study, to assess denoising quality and smoothness at the same time, we use the following goodness of denoising criterion:
\begin{equation}\label{snrrough}
\text{Goodness of denoising}(\hat{\mathbf{f}}) =10 \times \log_{10} \left(\text{SNR}(\hat{\mathbf{f}}) + \frac{1}{\text{Roughness}_N(\hat{\mathbf{f}})}\right)
\end{equation}
Equation \eqref{snrrough} refers to the decibel scale, which is a common scale in signal processing. When comparing two preprocessing methods on real data, we favor the method with a larger value of the aforementioned goodness of denoising criterion. 

Table \ref{tab:table5} displays the mean values (over all variables in the respective dataset indicated in the first column) of the criterion from \eqref{snrrough} for comparing the goodness of denoising of different methods. When applying SSA, we used the window length $L=N/2$. As it can be seen from Table \ref{tab:table5}, SSA achieved the largest average goodness of denoising value for all six considered datasets. According to our definition, SSA thus exhibits the best (average) denoising performance among all considered methods for all considered datasets.
\begin{table}[h]
\begin{center}
\begin{minipage}{220pt}
\caption{Results of assessing the goodness of denoising criterion for six noisy datasets. The displayed values refer to the decibel (dB) scale.}\label{tab:table5}%
\begin{tabular}{@{}lllll@{}}
\toprule		
			\textbf{Dataset} & \textbf{Basic SSA}& \textbf{ARFIMA} & \textbf{ETS}& \textbf{NN} \\ \hline
			AIS & 63.91   & 61.97 &  62.05 &  64.71 \\\hline
			Corona &  84.90& 84.55& 84.56& 84.59 \\\hline
			GDP &  81.55  &  66.51  &  60.16&59.91 \\\hline
			Import & 76.86&   59.28 &   59.07 &   60.46\\\hline
			Stuttgart &  115.55& 75.32& 75.66& 76.18\\\hline
			Halle &  117.03& 75.10& 75.62& 76.16 \\
\botrule
\end{tabular}
\end{minipage}
\end{center}
\end{table}

As a second criterion for assessing the quality of the decomposition of the observed data into a signal and a noise component, we considered the concept of separability. In fact, the SSA decomposition of the series $\textbf{Y}_N$  can only be successful if the signal and the noise components of the series are approximately separable from each other. Now, suppose that  based on the diagonal averaging step of SSA,  $ \widetilde{\textbf{X}}$ and 
$\widetilde{\textbf{Y}}$ are the trajectory matrices of two reconstructed series 
$\textbf{X}_N$  and $\textbf{Y}_N$, respectively. Define the $\textbf{w}$-scalar product of the series $\textbf{X}_N$  and $\textbf{Y}_N$ as $(\textbf{X}_N ,\textbf{Y}_N )_\textbf{w}=\sum_{k=1}^{N}w_kx_ky_k={\langle  \widetilde{\textbf{X}} ,  \widetilde{\textbf{Y}} \rangle}_F$, where ${\langle \cdot, \cdot \rangle}_F$ is the Frobenius inner product. Define the so-called $\textbf{w}$-correlation or weighted correlation as
\begin{equation}
\rho^{({\textbf{w}})}(\textbf{X}_N, \textbf{Y}_N) =\frac{\sum_{k=1}^{N}w_kx_ky_k}{\sqrt{\sum_{k=1}^{N}w_k x_k^{2}}\sqrt{\sum_{k=1}^{N} w_k y_k^{2}}},
\end{equation}
where $w_k$ is the frequency with which the series element $x_k$ appears in the related trajectory matrix, and it is given by $w_k = \min\{k, L, N-k+1\}$. If the absolute value of the $\textbf{w}$-correlation is close to zero, then the corresponding series are almost
$\textbf{w}$-orthogonal. If it is large, then the two series are far from being
$\textbf{w}$-orthogonal and are termed weakly separable. For more details, see Section 1.3 in \cite{Golyandina-R-2018}. 

Table \ref{tab.sep} displays the values of the $\textbf{w}$-correlation between the reconstructed and residuals vectors for the variables considered in Figure \ref{noisy data}. All $\textbf{w}$-correlation values in Table \ref{tab.sep} are rather small, indicating a good separation between the respective reconstructed and residuals vectors. 
\begin{table}[htp]
\begin{center}
\begin{minipage}{125pt}
\caption{Similarity between noise and extracted signal in terms of $\textbf{w}$-correlation}\label{tab.sep}
\begin{tabular}{@{}ll@{}}
\toprule
Variables &$\rho^{({\textbf{w}})}$\\
\hline
Red blood cell count & 5.82e-7\\ 
Hemoglobin level & 3.52e-7\\  
Gross fixed capital &  -1.82e-5\\  
Import price &  6.89e-4\\
Temperature  &  1.29e-8\\  
Divergence of wind &  8.04e-5\\  
\botrule
\end{tabular}
\end{minipage}
\end{center}
\end{table}

From the results presented in Tables  \ref{tab:table5} and Table \ref{tab.sep} it can be concluded that the SSA method is, among the considered preprocessing methods, the best choice for denoising the considered datasets.

\subsection{Multiple Testing Results}\label{mt.results} 
In this section, we apply the following two-step data analysis pipeline to the datasets that we have briefly presented in Section \ref{section.overview.data}.
\begin{description}
  \item[\textbf{Step 1:}] Application of the considered preprocessing methods in order to reduce the noise of the series
 \item[\textbf{Step 2:}] Application of appropriate multiple testing methods to identify variables with statistically significant effects / group differences
 \noindent
 \end{description}
In all following examples, we consider as the type I error criterion FWER control at level $\alpha=0.05$. 

\subsubsection{Australian Institute of Sport Data}\label{sport.data}
Here, we consider data of 102 male and 100 female athletes on 11 variables collected at the Australian Institute of Sport (AIS). This dataset is available in the \texttt{R} package $\texttt{sn}$. These data have been made publicly available in connection with the book by Cook and Weisberg (1994); see \cite{ais}. The focus of this study was to detect mean differences in the measured variables in female compared to male.  Denoting the sex-specific population means for variable $i$ by $\mu^{(i)}_{\text{male}}$ and $\mu^{(i)}_{\text{female}}$, respectively, the family of null hypotheses to be tested simultaneously is therefore given by
\begin{equation}\label{hyph}
\mathcal{H}_{11}=\{ H_{i}:1 \leq i \leq 11 \}, \quad \text{where} \quad H_{i}=\{ \mu^{(i)}_{\text{male}}=\mu^{(i)}_{\text{female}} \}.
\end{equation}
\noindent
As test statistics,  the $t$-statistics 
\begin{equation}\label{2ttest}
T_{i} = \frac{ \bar{Y}_{i,1}- \bar{Y}_{i,2}}{S_{P} \sqrt{ \frac{n_1+n_2}{n_1n_2}}} \quad \text{with} \quad S_{p}^{2}= \frac{(n_1-1)S_{i,1}^{2}+(n_2-1)S_{i,2}^{2}}{n_1+n_2-2}
\end{equation}
for $i \in \{1, \ldots, 11\}$ have been used, where $\bar{Y}_{i,1}, \bar{Y}_{i,2}$ and $S_{i,1}^{2}, S_{i,2}^{2}$ denote the empirical means and the empirical variances of males and females for variable $i$, and $n_1$ and $n_2$ denote the group-specific sample sizes. Letting $n=n_1 + n_2$, Student's $t$-distribution with $n-2$ degrees of freedom (with corresponding cumulative distribution function denoted by $F_{t_{n-2}}$) has been used as the null distribution of each of the test statistics, leading to the two-sided marginal $p$-value $p_{i}=2\left[1-F_{t_{n-2}}(\lvert T_{i} \rvert)\right]$ for variable $i \in \{1, \ldots, 11\}$. For performing the data analysis, the \texttt{R} packages \texttt{Rssa, multtest} and \texttt{mutoss} have been used.
Given that no assumption about the dependency structure among the marginal $p$-values is imposed, we applied the Bonferroni correction as well as Holm's procedure, which are controlling the FWER under any dependency structure; see Section \ref{multiple-tests-list}. For both of the latter multiple tests and for all considered preprocessing methods, we obtained $R_{11} = 10$ rejections.

\subsubsection{Coronavirus Data including Missing Values}\label{corona.data}
In this example, we consider several parameters with respect to their associations with the Coronavirus infection status, where the underlying data include missing values. The data have been collected from Imam Reza Hospital, Mashhad, Iran (\url{https://emamreza.mums.ac.ir/}). Here, in a first step of data analysis we imputed missing values by utilizing SSA, and the two-step data analysis pipeline given at the beginning of this section has then been applied to the imputed data. For imputing missing values, they are in a first step replaced by zeros,  and then updated in repeating cycles using basic SSA filtering until convergence is reached; see \cite{usr}. In total, $m=13$ parameters have been assessed for two independent samples, consisting of 1302 patients and 9761 healthy controls, respectively, which have been ordered based on the age of the subjects. 

In analogy to Section \ref{sport.data}, we have tested the $m=13$ null hypotheses of zero population mean differences (case group minus control group) corresponding to the $13$ measured parameters by calculating $13$ marginal $t$-statistics and their corresponding (two-sided) $p$-values.  Furthermore, since the measured parameters have to be assumed to be interrelated without further information about the type of resulting dependence structure among the $p$-values, we again applied the Bonferroni correction as well as Holm's procedure.  In this example, both considered multiple tests, in combination with any of the considered preprocessing methods, rejected all $13$ null hypotheses.

\subsubsection{Gross Domestic Product Data}\label{GDP.data}
The data that are used in this study are related to the UK and are taken from the Office for National Statistics (ONS).
In this example, $m=173$ different variables which are related to the Gross Domestic Product (GDP) are considered. The considered sampling period begins in the first quarter of 2001 and ends in the second quarter of 2021. The data have been divided into four year groups, namely 2001-2005, 2006-2010, 2011-2015, and 2016-2021. The goal of the statistical analysis is to find variables (indicators related to GDP) the population means of which differ between the four groups. Thus, the family of null hypotheses of interest is given by $\mathcal{H}_{173}=\{ H_i:1 \leq i \leq 173 \}$, where
\begin{equation}\label{4hyph}
H_{i}=\{ \mu^{(i)}_{\text{group}_{1}}=  \mu^{(i)}_{\text{group}_{2}}=\mu^{(i)}_{\text{group}_{3}}=\mu^{(i)}_{\text{group}_{4}}\}.
\end{equation}
In \eqref{4hyph}, $\mu^{(i)}_{\text{group}_{j}}$ denotes the population mean of variable $i$ in year group $j$.
For testing $\mathcal{H}_{173}$, we consider $F$-statistics of the form  
 \begin{equation*}
 F_{i}=\frac{\text{empirical between groups variance of variable~} i}{\text{empirical within  groups  variance of variable~} i}, \; 1 \leq i \leq 173.  
 \end{equation*}
Denoting the available sample size for variable $i$ by $N_i$, we assume as the null distribution of $F_i$ Fisher's $F$-distribution with $3$ and $N_{i}-3$ degrees of freedom, $1 \leq i \leq 173$, and this null distribution has been used to compute the corresponding $p$-value $p_i$.

Assuming positive dependence among the $p_i$'s, the Hochberg procedure as well as the single-step and the step-down version of the \v{S}id\'{a}k multiple test are appropriate for testing $\mathcal{H}_{173}$; see Section \ref{multiple-tests-list}. Table \ref{tab.gdp} tabulates the numbers of rejected null hypotheses for these three different multiple tests in connection with different preprocessing methods. In Table \ref{tab.gdp}, the numbers of rejected  null hypotheses for MSSA and ETS are close to each other. However, they are approximately $15\%$ and $20\%$ larger than those obtained by applying ARFIMA and NN, respectively.
\begin{table}[htp]
\begin{center}
\begin{minipage}{220pt}
\caption{Number of rejected null hypotheses based on different multiple tests and preprocessing methods for the dataset considered in Section \ref{GDP.data}.}\label{tab.gdp}%
\begin{tabular}{@{}llll@{}}
\toprule
Method & Hochberg& \v{S}id\'{a}k single-step& \v{S}id\'{a}k step-down \\\hline
MSSA &  154  &  146  &  154     \\\hline
ARFIMA &  135  &  131  &  135     \\\hline
ETS &  162  &  160  &  162     \\\hline
NN  &  126  &  123  &  126     \\
\botrule
\end{tabular}
\end{minipage}
\end{center}
\end{table}

\subsubsection{Import and Export Price Indices}\label{ie.data}
The data that are used in this example are related to Germany and are taken from GENESIS-Online Statistisches Bundesamt. The considered dataset includes $m = 76$ major components of import and export price indices. In all corresponding $76$ time series, the monthly sample period starts in January 2021 and ends in December 2021. The goal of the statistical analysis is to detect those components for which there is a difference between the population means in the import price index and the export price index. To this end, we analyzed each component $i$ by computing a $t$-statistic $T_i$ and the corresponding two-sided $p$-value $p_i$, where $i \in \{1, \ldots, 76\}$, in analogy to Sections \ref{sport.data} and \ref{corona.data}.

As in Section \ref{GDP.data}, we assumed positive dependence among the $p_i$'s and applied the Hochberg procedure as well as the single-step and the step-down version of the \v{S}id\'{a}k multiple test. In analogy to Table \ref{tab.gdp}, Table \ref{tab.ie} tabulates the obtained numbers of rejected null hypotheses for these three different multiple tests in connection with different preprocessing methods. In Table \ref{tab.ie}, the rejection numbers for MSSA are always larger than the rejection numbers for the other considered preprocessing methods.
\begin{table}[htp]
\begin{center}
\begin{minipage}{220pt}
\caption{Number of rejected null hypotheses based on different multiple tests and preprocessing methods for the dataset considered in Section \ref{ie.data}.}\label{tab.ie}%
\begin{tabular}{@{}lllll@{}}
\toprule
Method & Hochberg& \v{S}id\'{a}k single-step& \v{S}id\'{a}k step-down\\\hline
MSSA &  42  &  40  &  42     \\\hline
ARFIMA  &  38 &  35  &  38     \\\hline
ETS  &  37  &  34  &  37    \\\hline
NN &  36  &  35  &  36     \\
\botrule
\end{tabular}
\end{minipage}
\end{center}
\end{table}

\subsubsection{Climate Reanalyses Data from 2000 to Present}\label{weather.data}
Here we consider several atmospheric variables from the ERA-5 reanalysis dataset (see \cite{climate}) for the two German cities Halle and Stuttgart, from 2000 to present. These reanalyses combine observations and global climate model simulations with the goal of providing an accurate estimation of the state of the atmosphere at $0.25^\circ$ of horizontal resolution and on a global scale. The use of global climate simulations makes it possible for ERA-5 to cover gaps in observations over the globe, but it also introduces some uncertainty associated with the default values of the climate model. There are also uncertainties arising from the observations (they are limited in space and time and there are human and instrumental errors), and their number and precision have evolved over the years. 

The goal of the statistical analysis of this dataset is to identify those climate variables from $m=18$ variables for which the population mean referring to Halle differs from the population mean referring to Stuttgart. To this end, in analogy to Sections \ref{sport.data}, \ref{corona.data}, and \ref{ie.data}, we analyzed each climate variable $i$ by computing a $t$-statistic $T_i$ and the corresponding two-sided $p$-value $p_i$, where $i \in \{1, \ldots, 18\}$. Since we do not have dependence information in this example, we applied the Bonferroni test as well as Holm's procedure, because they are applicable under any kind of dependence structure among the $p$-values. In the case of applying the MSSA preprocessing method, we obtained $R_{18}(\varphi^{\text{Bonferroni}})=R_{18}(\varphi^{\text{Holm}})=15$ rejections. In the case of applying the ARFIMA preprocessing method, we obtained  $R_{18}(\varphi^{\text{Bonferroni}})=R_{18}(\varphi^{\text{Holm}})=11$ rejections, and for the ETS method as well as for the NN method, we obtained  $R_{18}(\varphi^{\text{Bonferroni}})=R_{18}(\varphi^{\text{Holm}})=10$ rejections. Thus, by applying MSSA as a preprocessing method, we obtained four or five additional rejections in comparison with the other considered preprocessing methods.

\subsubsection{Summary}
The main conclusion from all considered multiple test problems in this section is that appropriate data preprocessing can help to sharpen the statistical inference in the sense that a better denoising can lead to an increased statistical power of a multiple test which is applied on the denoised data series. (As mentioned in Section \ref{section.mt}, we take a larger number of rejected null hypotheses as an indication for higher statistical power when comparing the performance of two multiple tests, which control the FWER at the same level, on real data.) Since demonstrating this is the main purpose of this section, we abstained from a detailed discussion of the domain science implications of our data analysis results.

\section{Discussion}\label{section.conclusion}
We have compared four different data preprocessing methods with respect to their performance in denoising, both on simulated and on real data. In the case of simulated data, ground truth is known, and we have quantified the denoising accuracy by calculating RRMSE and RMAE referring to the signal component of the data. In the case of real data, ground truth is unknown, and we have introduced a goodness of denoising criterion which combines the empirical SNR with a roughness penalty. In both cases, SSA yielded overall the best denoising results.

Furthermore, we have assessed the impact of proper denoising on the performance of a variety of standard multiple tests for different study designs, including models with two independent groups, and models with more than two groups. Here, MSSA yielded overall the best results in terms of a large number of rejections under the constraint of FWER control. 

There are several directions for follow-up research: First, it may be of interest to carry out computer simulations in order to assess the impact of proper denoising on the type I error behavior of multiple tests when ground truth is known. In this respect, we have implicitly assumed in Section \ref{mt.results}  that the $p$-values originating from the different preprocessing methods are valid in the sense of Equation (1) in \cite{dickhaus2}. Second, given the empirical evidence that (multivariate) SSA has favorable denoising properties, it may be worthwhile to develop a mathematical optimality theory which underpins this empirical evidence by theoretical arguments. Third, from the perspective of applications, it appears appealing to interpret the estimated signal components of the considered data series in their respective domain science context.

\section*{Declarations}
Nothing to declare.


\begin{thebibliography}{48}
\ifx \bisbn   \undefined \def \bisbn  #1{ISBN #1}\fi
\ifx \binits  \undefined \def \binits#1{#1}\fi
\ifx \bauthor  \undefined \def \bauthor#1{#1}\fi
\ifx \batitle  \undefined \def \batitle#1{#1}\fi
\ifx \bjtitle  \undefined \def \bjtitle#1{#1}\fi
\ifx \bvolume  \undefined \def \bvolume#1{\textbf{#1}}\fi
\ifx \byear  \undefined \def \byear#1{#1}\fi
\ifx \bissue  \undefined \def \bissue#1{#1}\fi
\ifx \bfpage  \undefined \def \bfpage#1{#1}\fi
\ifx \blpage  \undefined \def \blpage #1{#1}\fi
\ifx \burl  \undefined \def \burl#1{\textsf{#1}}\fi
\ifx \doiurl  \undefined \def \doiurl#1{\url{https://doi.org/#1}}\fi
\ifx \betal  \undefined \def \betal{\textit{et al.}}\fi
\ifx \binstitute  \undefined \def \binstitute#1{#1}\fi
\ifx \binstitutionaled  \undefined \def \binstitutionaled#1{#1}\fi
\ifx \bctitle  \undefined \def \bctitle#1{#1}\fi
\ifx \beditor  \undefined \def \beditor#1{#1}\fi
\ifx \bpublisher  \undefined \def \bpublisher#1{#1}\fi
\ifx \bbtitle  \undefined \def \bbtitle#1{#1}\fi
\ifx \bedition  \undefined \def \bedition#1{#1}\fi
\ifx \bseriesno  \undefined \def \bseriesno#1{#1}\fi
\ifx \blocation  \undefined \def \blocation#1{#1}\fi
\ifx \bsertitle  \undefined \def \bsertitle#1{#1}\fi
\ifx \bsnm \undefined \def \bsnm#1{#1}\fi
\ifx \bsuffix \undefined \def \bsuffix#1{#1}\fi
\ifx \bparticle \undefined \def \bparticle#1{#1}\fi
\ifx \barticle \undefined \def \barticle#1{#1}\fi
\bibcommenthead
\ifx \bconfdate \undefined \def \bconfdate #1{#1}\fi
\ifx \botherref \undefined \def \botherref #1{#1}\fi
\ifx \url \undefined \def \url#1{\textsf{#1}}\fi
\ifx \bchapter \undefined \def \bchapter#1{#1}\fi
\ifx \bbook \undefined \def \bbook#1{#1}\fi
\ifx \bcomment \undefined \def \bcomment#1{#1}\fi
\ifx \oauthor \undefined \def \oauthor#1{#1}\fi
\ifx \citeauthoryear \undefined \def \citeauthoryear#1{#1}\fi
\ifx \endbibitem  \undefined \def \endbibitem {}\fi
\ifx \bconflocation  \undefined \def \bconflocation#1{#1}\fi
\ifx \arxivurl  \undefined \def \arxivurl#1{\textsf{#1}}\fi
\csname PreBibitemsHook\endcsname

\bibitem{Pyle1999}
\begin{bbook}
\bauthor{\bsnm{Pyle}, \binits{D.}}:
\bbtitle{Data Preparation for Data Mining}.
\bpublisher{Morgan Kaufmann Publishers},
\blocation{Los Altos, California}
(\byear{1999})
\end{bbook}
\endbibitem

\bibitem{Repsilber2008}
\begin{bchapter}
\bauthor{\bsnm{Repsilber}, \binits{D.}}:
\bctitle{From spots to candidates: image analysis, data processing, candidate
  selection}.
In: \bbtitle{H{\"{a}}mostaseologie},
vol. \bseriesno{28, A5 WS-04-02-}
(\byear{2008})
\end{bchapter}
\endbibitem

\bibitem{OLIVERI20199}
\begin{barticle}
\bauthor{\bsnm{Oliveri}, \binits{P.}},
\bauthor{\bsnm{Malegori}, \binits{C.}},
\bauthor{\bsnm{Simonetti}, \binits{R.}},
\bauthor{\bsnm{Casale}, \binits{M.}}:
\batitle{The impact of signal pre-processing on the final interpretation of
  analytical outcomes – a tutorial}.
\bjtitle{Analytica Chimica Acta}
\bvolume{1058},
\bfpage{9}--\blpage{17}
(\byear{2019}).
\doiurl{10.1016/j.aca.2018.10.055}
\end{barticle}
\endbibitem

\bibitem{Hyndman20081}
\begin{barticle}
\bauthor{\bsnm{Hyndman}, \binits{R.J.}},
\bauthor{\bsnm{Khandakar}, \binits{Y.}}:
\batitle{Automatic time series forecasting: The forecast package for {R}}.
\bjtitle{Journal of Statistical Software}
\bvolume{27}(\bissue{3}),
\bfpage{1}--\blpage{22}
(\byear{2008}).
\doiurl{10.18637/jss.v027.i03}
\end{barticle}
\endbibitem

\bibitem{b1}
\begin{bbook}
\bauthor{\bsnm{Hyndman}, \binits{R.}},
\bauthor{\bsnm{Athanasopoulos}, \binits{G.}}:
\bbtitle{Forecasting: Principles and Practice},
\bedition{3rd} edn.
\bpublisher{OTexts},
\blocation{Melbourne, Australia}
(\byear{2021})
\end{bbook}
\endbibitem

\bibitem{Bishop-NN}
\begin{bbook}
\bauthor{\bsnm{Bishop}, \binits{C.M.}}:
\bbtitle{Neural Networks for Pattern Recognition}.
\bpublisher{The Clarendon Press, Oxford University Press},
\blocation{New York, NY}
(\byear{1995}).
\bcomment{With a foreword by Geoffrey Hinton}
\end{bbook}
\endbibitem

\bibitem{Dickhaus2018}
\begin{bbook}
\bauthor{\bsnm{Dickhaus}, \binits{T.}}:
\bbtitle{Theory of Nonparametric Tests}.
\bpublisher{Springer},
\blocation{Cham, Switzerland}
(\byear{2018}).
\burl{https://doi.org/10.1007/978-3-319-76315-6}
\end{bbook}
\endbibitem

\bibitem{Hassani2013}
\begin{barticle}
\bauthor{\bsnm{Hassani}, \binits{H.}},
\bauthor{\bsnm{Soofi}, \binits{A.S.}},
\bauthor{\bsnm{Zhigljavsky}, \binits{A.}}:
\batitle{Predicting inflation dynamics with singular spectrum analysis}.
\bjtitle{Journal of the Royal Statistical Society. Series A. Statistics in
  Society}
\bvolume{176}(\bissue{3}),
\bfpage{743}--\blpage{760}
(\byear{2013}).
\doiurl{10.1111/j.1467-985X.2012.01061.x}
\end{barticle}
\endbibitem

\bibitem{Handbook}
\begin{bbook}
\bauthor{\bsnm{{Cui, Xinping and Dickhaus, Thorsten and Ding, Ying and Hsu,
  Jason C. (Eds.)}}}:
\bbtitle{Handbook of Multiple Comparisons}.
\bsertitle{{Chapman \& Hall/CRC Handbooks of Modern Statistical Methods}}.
\bpublisher{CRC Press},
\blocation{Boca Raton, FL}
(\byear{2022})
\end{bbook}
\endbibitem

\bibitem{thorsten}
\begin{bbook}
\bauthor{\bsnm{Dickhaus}, \binits{T.}}:
\bbtitle{Simultaneous Statistical Inference with Applications in the Life
  Sciences}.
\bpublisher{Springer},
\blocation{Berlin, Heidelberg}
(\byear{2014}).
\burl{https://doi.org/10.1007/978-3-642-45182-9}
\end{bbook}
\endbibitem

\bibitem{BROOMHEAD1986217}
\begin{barticle}
\bauthor{\bsnm{Broomhead}, \binits{D.S.}},
\bauthor{\bsnm{King}, \binits{G.P.}}:
\batitle{Extracting qualitative dynamics from experimental data}.
\bjtitle{Physica D: Nonlinear Phenomena}
\bvolume{20}(\bissue{2-3}),
\bfpage{217}--\blpage{236}
(\byear{1986}).
\doiurl{10.1016/0167-2789(86)90031-X}
\end{barticle}
\endbibitem

\bibitem{MOVAHEDIFAR201852}
\begin{barticle}
\bauthor{\bsnm{Movahedifar}, \binits{M.}},
\bauthor{\bsnm{Yarmohammadi}, \binits{M.}},
\bauthor{\bsnm{Hassani}, \binits{H.}}:
\batitle{Bicoid signal extraction: another powerful approach}.
\bjtitle{Mathematical Biosciences}
\bvolume{303},
\bfpage{52}--\blpage{61}
(\byear{2018}).
\doiurl{10.1016/j.mbs.2018.06.002}
\end{barticle}
\endbibitem

\bibitem{SILVA2019134}
\begin{barticle}
\bauthor{\bsnm{Silva}, \binits{E.S.}},
\bauthor{\bsnm{Hassani}, \binits{H.}},
\bauthor{\bsnm{Heravi}, \binits{S.}},
\bauthor{\bsnm{Huang}, \binits{X.}}:
\batitle{Forecasting tourism demand with denoised neural networks}.
\bjtitle{Annals of Tourism Research}
\bvolume{74},
\bfpage{134}--\blpage{154}
(\byear{2019}).
\doiurl{10.1016/j.annals.2018.11.006}
\end{barticle}
\endbibitem

\bibitem{article35}
\begin{barticle}
\bauthor{\bsnm{Silva}, \binits{E.S.}},
\bauthor{\bsnm{Hassani}, \binits{H.}},
\bauthor{\bsnm{Heravi}, \binits{S.}}:
\batitle{Modeling {E}uropean industrial production with multivariate singular
  spectrum analysis: a cross-industry analysis}.
\bjtitle{Journal of Forecasting}
\bvolume{37}(\bissue{3}),
\bfpage{371}--\blpage{384}
(\byear{2018}).
\doiurl{10.1002/for.2508}
\end{barticle}
\endbibitem

\bibitem{HASSANI38}
\begin{barticle}
\bauthor{\bsnm{Hassani}, \binits{H.}},
\bauthor{\bsnm{Silva}, \binits{E.S.}},
\bauthor{\bsnm{Gupta}, \binits{R.}},
\bauthor{\bsnm{Das}, \binits{S.}}:
\batitle{Predicting global temperature anomaly: a definitive investigation
  using an ensemble of twelve competing forecasting models}.
\bjtitle{Physica A. Statistical Mechanics and its Applications}
\bvolume{509},
\bfpage{121}--\blpage{139}
(\byear{2018}).
\doiurl{10.1016/j.physa.2018.05.147}
\end{barticle}
\endbibitem

\bibitem{article39}
\begin{barticle}
\bauthor{\bsnm{Hassani}, \binits{H.}},
\bauthor{\bsnm{Rua}, \binits{A.}},
\bauthor{\bsnm{Silva}, \binits{E.}},
\bauthor{\bsnm{Thomakos}, \binits{D.}}:
\batitle{Monthly forecasting of gdp with mixed-frequency multivariate singular
  spectrum analysis}.
\bjtitle{International Journal of Forecasting}
\bvolume{35},
\bfpage{1263}--\blpage{1272}
(\byear{2019}).
\doiurl{10.1016/j.ijforecast.2019.03.021}
\end{barticle}
\endbibitem

\bibitem{signals}
\begin{barticle}
\bauthor{\bsnm{Hassani}, \binits{H.}},
\bauthor{\bsnm{Yeganegi}, \binits{M.R.}},
\bauthor{\bsnm{Khan}, \binits{A.}},
\bauthor{\bsnm{Silva}, \binits{E.S.}}:
\batitle{The effect of data transformation on singular spectrum analysis for
  forecasting}.
\bjtitle{Signals}
\bvolume{1}(\bissue{1}),
\bfpage{4}--\blpage{25}
(\byear{2020}).
\doiurl{10.3390/signals1010002}
\end{barticle}
\endbibitem

\bibitem{SILVA2017}
\begin{barticle}
\bauthor{\bsnm{Silva}, \binits{E.S.}},
\bauthor{\bsnm{Ghodsi}, \binits{Z.}},
\bauthor{\bsnm{Ghodsi}, \binits{M.}},
\bauthor{\bsnm{Heravi}, \binits{S.}},
\bauthor{\bsnm{Hassani}, \binits{H.}}:
\batitle{Cross country relations in european tourist arrivals}.
\bjtitle{Annals of Tourism Research}
\bvolume{63},
\bfpage{151}--\blpage{168}
(\byear{2017}).
\doiurl{10.1016/j.annals.2017.01.012}
\end{barticle}
\endbibitem

\bibitem{DECARVALHO}
\begin{barticle}
\bauthor{\bsnm{{de Carvalho}}, \binits{M.}},
\bauthor{\bsnm{Rua}, \binits{A.}}:
\batitle{Real-time nowcasting the us output gap: Singular spectrum analysis at
  work}.
\bjtitle{International Journal of Forecasting}
\bvolume{33}(\bissue{1}),
\bfpage{185}--\blpage{198}
(\byear{2017}).
\doiurl{10.1016/j.ijforecast.2015.09.004}
\end{barticle}
\endbibitem

\bibitem{articlehs}
\begin{barticle}
\bauthor{\bsnm{Hassani}, \binits{H.}},
\bauthor{\bsnm{Silva}, \binits{E.}}:
\batitle{Forecasting energy data with a time lag into the future and google
  trends}.
\bjtitle{International Journal of Energy and Statistics}
\bvolume{4},
\bfpage{1650020}
(\byear{2016}).
\doiurl{10.1142/S2335680416500204}
\end{barticle}
\endbibitem

\bibitem{articlepl}
\begin{barticle}
\bauthor{\bsnm{Portes}, \binits{L.L.}},
\bauthor{\bsnm{Aguirre}, \binits{L.A.}}:
\batitle{Enhancing multivariate singular spectrum analysis for phase
  synchronization: the role of observability}.
\bjtitle{Chaos}
\bvolume{26}(\bissue{9}),
\bfpage{093112}--\blpage{12}
(\byear{2016}).
\doiurl{10.1063/1.4963013}
\end{barticle}
\endbibitem

\bibitem{Golyandina-2001}
\begin{bbook}
\bauthor{\bsnm{Golyandina}, \binits{N.}},
\bauthor{\bsnm{Nekrutkin}, \binits{V.}},
\bauthor{\bsnm{Zhigljavsky}, \binits{A.}}:
\bbtitle{Analysis of Time Series Structure: SSA and Related Techniques}.
\bsertitle{Monographs on Statistics and Applied Probability},
vol. \bseriesno{90}.
\bpublisher{Chapman \& Hall/CRC},
\blocation{Boca Raton, FL}
(\byear{2001}).
\burl{https://doi.org/10.1201/9781420035841}
\end{bbook}
\endbibitem

\bibitem{Golyandina-Zhigljavsky-2013}
\begin{bbook}
\bauthor{\bsnm{Golyandina}, \binits{N.}},
\bauthor{\bsnm{Zhigljavsky}, \binits{A.}}:
\bbtitle{Singular Spectrum Analysis for Time Series}.
\bpublisher{Springer},
\blocation{Berlin, Heidelberg}
(\byear{2013}).
\burl{https://doi.org/10.1007/978-3-642-34913-3}
\end{bbook}
\endbibitem

\bibitem{Golyandina-R-2018}
\begin{bbook}
\bauthor{\bsnm{Golyandina}, \binits{N.}},
\bauthor{\bsnm{Korobeynikov}, \binits{A.}},
\bauthor{\bsnm{Zhigljavsky}, \binits{A.}}:
\bbtitle{Singular Spectrum Analysis with {R}}.
\bpublisher{Springer},
\blocation{Berlin}
(\byear{2018}).
\burl{https://doi.org/10.1007/978-3-662-57380-8}
\end{bbook}
\endbibitem

\bibitem{arfima}
\begin{barticle}
\bauthor{\bsnm{Bannor}, \binits{R.}},
\bauthor{\bsnm{Mada}, \binits{M.}}:
\batitle{{Forecasting wholesale price of cluster bean using the Autoregressive
  Fractionally Integrated Moving-Average Model: the case of Sri Ganganagar of
  Rajasthan in India}}.
\bjtitle{Journal of Business Management and Economics}
\bvolume{3},
\bfpage{1}--\blpage{7}
(\byear{2015}).
\doiurl{10.15520/jbme.2015.vol3.iss8.132.pp01-07}
\end{barticle}
\endbibitem

\bibitem{arfima2}
\begin{barticle}
\bauthor{\bsnm{Dhliwayo}, \binits{L.}},
\bauthor{\bsnm{Matarise}, \binits{F.}},
\bauthor{\bsnm{Chimedza}, \binits{C.}}:
\batitle{Autoregressive fractionally integrated moving average-generalized
  autoregressive conditional heteroskedasticity model with level shift
  intervention}.
\bjtitle{Open Journal of Statistics}
\bvolume{10},
\bfpage{341}--\blpage{362}
(\byear{2020}).
\doiurl{10.4236/ojs.2020.102023}
\end{barticle}
\endbibitem

\bibitem{HYNDMAN2002439}
\begin{barticle}
\bauthor{\bsnm{Hyndman}, \binits{R.J.}},
\bauthor{\bsnm{Koehler}, \binits{A.B.}},
\bauthor{\bsnm{Snyder}, \binits{R.D.}},
\bauthor{\bsnm{Grose}, \binits{S.}}:
\batitle{A state space framework for automatic forecasting using exponential
  smoothing methods}.
\bjtitle{International Journal of Forecasting}
\bvolume{18}(\bissue{3}),
\bfpage{439}--\blpage{454}
(\byear{2002}).
\doiurl{10.1016/S0169-2070(01)00110-8}
\end{barticle}
\endbibitem

\bibitem{Hyndman2008}
\begin{bbook}
\bauthor{\bsnm{{Hyndman}}, \binits{R.J.}},
\bauthor{\bsnm{{Koehler}}, \binits{A.B.}},
\bauthor{\bsnm{{Ord}}, \binits{J.K.}},
\bauthor{\bsnm{{Snyder}}, \binits{R.D.}}:
\bbtitle{Forecasting with Exponential Smoothing. The State Space Approach}.
\bsertitle{{Springer Series in Statistics}}.
\bpublisher{Springer},
\blocation{Berlin}
(\byear{2008})
\end{bbook}
\endbibitem

\bibitem{Philipp-ICLR2017}
\begin{bchapter}
\bauthor{\bsnm{Philipp}, \binits{G.}},
\bauthor{\bsnm{Carbonell}, \binits{J.G.}}:
\bctitle{Nonparametric neural networks}.
In: \bbtitle{Proceedings of the International Conference on Learning
  Representations}
(\byear{2017})
\end{bchapter}
\endbibitem

\bibitem{Bonferroni1}
\begin{botherref}
\oauthor{\bsnm{Bonferroni}, \binits{C.E.}}:
Il calcolo delle assicurazioni su gruppi di teste.
Studi in onore del Professore Salvatore Ortu Carboni,
13--60
(1935)
\end{botherref}
\endbibitem

\bibitem{Bonferroni2}
\begin{barticle}
\bauthor{\bsnm{Bonferroni}, \binits{C.E.}}:
\batitle{Teoria statistica delle classi e calcolo delle probabilita}.
\bjtitle{Pubblicazioni del R Istituto Superiore di Scienze Economiche e
  Commericiali di Firenze}
\bvolume{8},
\bfpage{3}--\blpage{62}
(\byear{1936})
\end{barticle}
\endbibitem

\bibitem{holm}
\begin{barticle}
\bauthor{\bsnm{Holm}, \binits{S.}}:
\batitle{A simple sequentially rejective multiple test procedure}.
\bjtitle{Scandinavian Journal of Statistics}
\bvolume{6}(\bissue{2}),
\bfpage{65}--\blpage{70}
(\byear{1979})
\end{barticle}
\endbibitem

\bibitem{sidak1967}
\begin{barticle}
\bauthor{\bsnm{{\v{S}}id{\'{a}}k}, \binits{Z.}}:
\batitle{{Rectangular confidence regions for the means of multivariate normal
  distributions.}}
\bjtitle{J. Am. Stat. Assoc.}
\bvolume{62},
\bfpage{626}--\blpage{633}
(\byear{1967}).
\doiurl{10.2307/2283989}
\end{barticle}
\endbibitem

\bibitem{MPHTP}
\begin{barticle}
\bauthor{\bsnm{Dickhaus}, \binits{T.}},
\bauthor{\bsnm{Stange}, \binits{J.}}:
\batitle{{Multiple point hypothesis test problems and effective numbers of
  tests for control of the family-wise error rate}}.
\bjtitle{Calcutta Statistical Association Bulletin}
\bvolume{65}(\bissue{257-260}),
\bfpage{123}--\blpage{144}
(\byear{2013})
\end{barticle}
\endbibitem

\bibitem{hoch}
\begin{barticle}
\bauthor{\bsnm{Hochberg}, \binits{Y.}}:
\batitle{A sharper {B}onferroni procedure for multiple tests of significance}.
\bjtitle{Biometrika}
\bvolume{75}(\bissue{4}),
\bfpage{800}--\blpage{802}
(\byear{1988}).
\doiurl{10.1093/biomet/75.4.800}
\end{barticle}
\endbibitem

\bibitem{sim}
\begin{barticle}
\bauthor{\bsnm{Simes}, \binits{R.J.}}:
\batitle{An improved {B}onferroni procedure for multiple tests of
  significance}.
\bjtitle{Biometrika}
\bvolume{73}(\bissue{3}),
\bfpage{751}--\blpage{754}
(\byear{1986}).
\doiurl{10.1093/biomet/73.3.751}
\end{barticle}
\endbibitem

\bibitem{Bodnar-Dickhaus-AISM}
\begin{barticle}
\bauthor{\bsnm{Bodnar}, \binits{T.}},
\bauthor{\bsnm{Dickhaus}, \binits{T.}}:
\batitle{On the {S}imes inequality in elliptical models}.
\bjtitle{Ann. Inst. Statist. Math.}
\bvolume{69}(\bissue{1}),
\bfpage{215}--\blpage{230}
(\byear{2017}).
\doiurl{10.1007/s10463-015-0539-4}
\end{barticle}
\endbibitem

\bibitem{Finner-Simes}
\begin{barticle}
\bauthor{\bsnm{Finner}, \binits{H.}},
\bauthor{\bsnm{Roters}, \binits{M.}},
\bauthor{\bsnm{Strassburger}, \binits{K.}}:
\batitle{On the {S}imes test under dependence}.
\bjtitle{Statist. Papers}
\bvolume{58}(\bissue{3}),
\bfpage{775}--\blpage{789}
(\byear{2017}).
\doiurl{10.1007/s00362-015-0725-8}
\end{barticle}
\endbibitem

\bibitem{thor1}
\begin{barticle}
\bauthor{\bsnm{Schildknecht}, \binits{K.}},
\bauthor{\bsnm{Olek}, \binits{S.}},
\bauthor{\bsnm{Dickhaus}, \binits{T.}}:
\batitle{Simultaneous statistical inference for epigenetic data}.
\bjtitle{PLOS ONE}
\bvolume{10},
\bfpage{0125587}
(\byear{2015}).
\doiurl{10.1371/journal.pone.0125587}
\end{barticle}
\endbibitem

\bibitem{thor2}
\begin{barticle}
\bauthor{\bsnm{Neumann}, \binits{A.}},
\bauthor{\bsnm{Bodnar}, \binits{T.}},
\bauthor{\bsnm{Pfeifer}, \binits{D.}},
\bauthor{\bsnm{Dickhaus}, \binits{T.}}:
\batitle{Multivariate multiple test procedures based on nonparametric copula
  estimation}.
\bjtitle{Biometrical Journal}
\bvolume{61}(\bissue{1}),
\bfpage{40}--\blpage{61}
(\byear{2019}).
\doiurl{10.1002/bimj.201700205}
\end{barticle}
\endbibitem

\bibitem{thor3}
\begin{barticle}
\bauthor{\bsnm{Dickhaus}, \binits{T.}}:
\batitle{Simultaneous {B}ayesian analysis of contingency tables in genetic
  association studies}.
\bjtitle{Statistical Applications in Genetics and Molecular Biology}
\bvolume{14}(\bissue{4}),
\bfpage{347}--\blpage{360}
(\byear{2015}).
\doiurl{10.1515/sagmb-2014-0052}
\end{barticle}
\endbibitem

\bibitem{thor4}
\begin{bchapter}
\bauthor{\bsnm{Dickhaus}, \binits{T.}}:
\bctitle{Combining high-dimensional classification and multiple hypotheses
  testing for the analysis of big data in genetics}.
In: \bbtitle{Statistics and Its Applications}.
\bsertitle{Springer Proc. Math. Stat.},
vol. \bseriesno{244},
pp. \bfpage{47}--\blpage{50}.
\bpublisher{Springer},
\blocation{Singapore}
(\byear{2018}).
\doiurl{10.1007/978-981-13-1223-6_5}
\end{bchapter}
\endbibitem

\bibitem{dickhaus2}
\begin{barticle}
\bauthor{\bsnm{Hoang}, \binits{A.-T.}},
\bauthor{\bsnm{Dickhaus}, \binits{T.}}:
\batitle{Randomized ‐values for multiple testing and their application in
  replicability analysis}.
\bjtitle{Biometrical Journal}
\bvolume{64}(\bissue{2}),
\bfpage{384}--\blpage{409}
(\byear{2021}).
\doiurl{10.1002/bimj.202000155}
\end{barticle}
\endbibitem

\bibitem{ghafari}
\begin{barticle}
\bauthor{\bsnm{Gorji}, \binits{F.}},
\bauthor{\bsnm{Aminghafari}, \binits{M.}}:
\batitle{Denoising heavy-tailed data using a novel robust non-parametric method
  based on quantile regression}.
\bjtitle{Fluctuation and Noise Letters}
\bvolume{16},
\bfpage{1750029}
(\byear{2017}).
\doiurl{10.1142/S0219477517500298}
\end{barticle}
\endbibitem

\bibitem{smooth2}
\begin{bbook}
\bauthor{\bsnm{Fahrmeir}, \binits{L.}},
\bauthor{\bsnm{Tutz}, \binits{G.}}:
\bbtitle{Multivariate Statistical Modelling Based on Generalized Linear Models.
  Second Edition}.
\bsertitle{Springer Series in Statistics}.
\bpublisher{Springer},
\blocation{New York}
(\byear{2001}).
\burl{https://doi.org/10.1007/978-1-4757-3454-6}
\end{bbook}
\endbibitem

\bibitem{ais}
\begin{bbook}
\bauthor{\bsnm{Cook}, \binits{R.D.}},
\bauthor{\bsnm{Weisberg}, \binits{S.}}:
\bbtitle{An Introduction to Regression Graphics}.
\bsertitle{Wiley Series in Probability and Statistics}.
\bpublisher{John Wiley and Sons Inc},
\blocation{New York, NY}
(\byear{1994})
\end{bbook}
\endbibitem

\bibitem{usr}
\begin{bbook}
\bauthor{\bsnm{Hassani}, \binits{H.}},
\bauthor{\bsnm{Mahmoudvand}, \binits{R.}}:
\bbtitle{Singular Spectrum Analysis Using R}.
\bpublisher{Palgrave Macmillan},
\blocation{Basingstoke, UK}
(\byear{2018})
\end{bbook}
\endbibitem

\bibitem{climate}
\begin{barticle}
\bauthor{\bsnm{Hersbach}, \binits{H.}},
\bauthor{\bsnm{Bell}, \binits{B.}},
\bauthor{\bsnm{Berrisford}, \binits{P.}},
\bauthor{\bsnm{Hirahara}, \binits{S.}},
\bauthor{\bsnm{Horányi}, \binits{A.}},
\bauthor{\bsnm{Muñoz-Sabater}, \binits{J.}},
\bauthor{\bsnm{Nicolas}, \binits{J.}},
\bauthor{\bsnm{Peubey}, \binits{C.}},
\bauthor{\bsnm{Radu}, \binits{R.}},
\bauthor{\bsnm{Schepers}, \binits{D.}},
\bauthor{\bsnm{Simmons}, \binits{A.}},
\bauthor{\bsnm{Soci}, \binits{C.}},
\bauthor{\bsnm{Abdalla}, \binits{S.}},
\bauthor{\bsnm{Abellan}, \binits{X.}},
\bauthor{\bsnm{Balsamo}, \binits{G.}},
\bauthor{\bsnm{Bechtold}, \binits{P.}},
\bauthor{\bsnm{Biavati}, \binits{G.}},
\bauthor{\bsnm{Bidlot}, \binits{J.}},
\bauthor{\bsnm{Bonavita}, \binits{M.}},
\bauthor{\bsnm{De~Chiara}, \binits{G.}},
\bauthor{\bsnm{Dahlgren}, \binits{P.}},
\bauthor{\bsnm{Dee}, \binits{D.}},
\bauthor{\bsnm{Diamantakis}, \binits{M.}},
\bauthor{\bsnm{Dragani}, \binits{R.}},
\bauthor{\bsnm{Flemming}, \binits{J.}},
\bauthor{\bsnm{Forbes}, \binits{R.}},
\bauthor{\bsnm{Fuentes}, \binits{M.}},
\bauthor{\bsnm{Geer}, \binits{A.}},
\bauthor{\bsnm{Haimberger}, \binits{L.}},
\bauthor{\bsnm{Healy}, \binits{S.}},
\bauthor{\bsnm{Hogan}, \binits{R.J.}},
\bauthor{\bsnm{Hólm}, \binits{E.}},
\bauthor{\bsnm{Janisková}, \binits{M.}},
\bauthor{\bsnm{Keeley}, \binits{S.}},
\bauthor{\bsnm{Laloyaux}, \binits{P.}},
\bauthor{\bsnm{Lopez}, \binits{P.}},
\bauthor{\bsnm{Lupu}, \binits{C.}},
\bauthor{\bsnm{Radnoti}, \binits{G.}},
\bauthor{\bparticle{de} \bsnm{Rosnay}, \binits{P.}},
\bauthor{\bsnm{Rozum}, \binits{I.}},
\bauthor{\bsnm{Vamborg}, \binits{F.}},
\bauthor{\bsnm{Villaume}, \binits{S.}},
\bauthor{\bsnm{Thépaut}, \binits{J.-N.}}:
\batitle{{The ERA5 global reanalysis}}.
\bjtitle{Quarterly Journal of the Royal Meteorological Society}
\bvolume{146}(\bissue{730}),
\bfpage{1999}--\blpage{2049}
(\byear{2020}).
\doiurl{10.1002/qj.3803}
\end{barticle}
\endbibitem

\end{thebibliography}


\end{document}